\documentstyle[prl,aps,preprint]{revtex}
\begin{document}
\draft
\title{Quantum Cryptography in Noisy Channels}
\author{Hoi-Kwong Lo\footnotemark ~and H. F. Chau\footnotemark}
\address{
 School of Natural Sciences, Institute for Advanced Study, Olden Lane,
 Princeton, NJ 08540
}
\date{\today}
\preprint{IASSNS-HEP-95/93}
\footnotetext[1]{e-mail: hkl@sns.ias.edu}
\footnotetext[2]{e-mail: chau@sns.ias.edu}
\maketitle
\begin{abstract}
 We provide a complete proof of the security of
 quantum cryptography against any eavesdropping
 attack including coherent measurements even in the presence
 of noise.
 Polarization-based cryptographic schemes are
 shown to be equivalent to EPR-based schemes.
 We also show that
 the performance of a noisy channel approaches that of a noiseless
 one as the error rate tends to zero.
 (i.e., the secrecy capacity
 $C_s (\epsilon) \to C_s (0)$ as $\epsilon \to 0$.)
 One implication of our results is that one can {\it double} the
 efficiency of a most well-known quantum cryptographic scheme
 proposed by Bennett and Brassard
 simply by assigning vastly different probabilities to the
 two conjugate bases.
\end{abstract}
\pacs{\noindent
 \begin{minipage}[t]{5.5in}\vspace{1ex}
  \vspace{1ex} \\
 \begin{tabular}{lll}
  AMS 1991 Subject Classification: & (Primary) & 68P25, 81P15, 94A17, 94A24,\\
 & & 94A60 \\
  & (Secondary)&  68P20, 81V99
 \end{tabular}
 \vspace{0.9ex} \\
 \begin{tabular}{ll}
  Keywords: & channel capacity, coherent measurement, quantum cryptography, \\
  & quantum information theory, secrecy capacity
 \end{tabular}
 \end{minipage}
}
\section{Introduction}
\label{sec:intro}
 Cryptography is the art of providing secure communication over insecure
 (i.e., subject to eavesdropping) communication channels. The security
 of a conventional cryptosystem often lies on a relatively short secret
 value known as the key that has to be agreed on by the two legitimate
 users before secure communication can be started.  For this reason,
 secure key distribution is a crucial issue in cryptography. Unfortunately,
 classical cryptography provides no tools to guarantee the security of the
 key distribution because classical signals are vulnerable to passive
 interception: A passive wiretapper can simply make copies (clones) of the
 carrier of information and read off from those copies the value of the key.
 Since the original carrier of information can be resent to the legitimate
 user without alteration, there is no way for the two users to check whether
 the carrier has been intercepted.

 In quantum mechanics, any measurement that does not disturb a
 complete set of non-orthogonal states also fails to yield any
 information distinguishing them \cite{Bennett:92}. (See Appendix for
 a proof.)
 In particular, it is impossible for the eavesdropper to
 clone non-orthogonal states.
 Therefore, coding based on non-orthogonal states can be used to
 detect any eavesdropping attempt \cite{Wiesner:83}.
 The feasibility of secure quantum
 key distribution over long distance
 by optical fiber has been recently demonstrated: A prototype system
 at BT laboratories is capable of key transfer over 10~km in optical
 fiber at date rates of 20~kbit\,s$^{-1}$ \cite{Franson:94}.
 The investigation
 on the foundations of quantum cryptography is thus timely.

 Noise is inevitable in any real communication channel.
 It is, therefore, crucial to demonstrate the security of quantum
 cryptography when the communication channel is noisy.
 Various eavesdropping strategies have been investigated in
 the literature \cite{Exper:92}. In order to acquire any appreciable amount of
 information about the transmitted signals, they are all shown to
 introduce a substantial change
 in the error rate. Therefore, quantum cryptography is generally
 {\it conjectured} to be secure. Unfortunately,
 it has not yet been ruled out that
 still more sophisticated use of
 quantum physics might defeat quantum cryptography\footnote{Deutsch
{\it et al.} \cite{Jozsa:un} have
 suggested a purification scheme in which the two legitimate
 users perform coherent manipulations on the transmitted
 particles. Such a scheme is asymptotically unconditionally safe against
 any attack. Unfortunately, its efficiency is
 very limited.}.
 This is hardly a comforting
 situation: In the long history of classical
 cryptography, there were numerous instances of unexpected
 failures of cryptographic schemes (e.g., the knapsack
 scheme) that were once believed to be unbreakable.
 Such failures often led to dramatic and even disastrous consequences.
 To ensure
 that quantum cryptography does not follow the same trail, it is,
 therefore,
 essential for us to establish rigorously its absolute security.
 The first goal of this paper is to give such a proof.

 The most general eavesdropping
 strategy available to an eavesdropper, traditionally called Eve, is for
 her to {\it coherently}
 manipulate all the transmitted particles \cite{Jozsa:un} by
 coupling them as a single entity with a probe (an ancilla).
 Eve may subsequently
 perform measurements on the ancilla to acquire information about
 the transmission.
 In Section~\ref{sec:coh}, we prove
 the security of quantum cryptography in a noisy channel
 by showing that it is unbreakable even by coherent manipulations
 performed by the eavesdropper.

 After establishing the security of quantum cryptography, we come
 to the next
 question: {\it how much} information can be securely transmitted
 through a noisy quantum communication channel?
 Both eavesdropping and the intrinsic noise of the system introduce
 errors (including decoherence) in the channel and it is
 often difficult to distinguish between the two sources.
 Therefore, a conservative user
 may assume that all the errors are due to the wiretapping.
 Since wiretapping in a quantum channel necessarily leads to
 errors in the transmission, the legitimate users can put an upper
 bound to the extent of wiretapping by determining the error rate
 of the channel. Standard techniques such as
 error-correcting codes and privacy amplification can then be applied
 to the partly secret raw signals
 to distill a shorter but absolutely secure sequence of bits
 which can then be used as the key for subsequent classical
 communication. An insecure but unjammable (i.e., subject to wiretapping but
 not alteration of messages) classical channel
 can be used for the public discussion between the two users
 during the distillation process.

 The maximal number of secure bits that can be distilled
 from each raw signal transmitted through a quantum channel is clearly
 a function of the error rate. It is defined as the secrecy
 capacity \cite{Maurer:93} of the quantum channel. The second
 goal of this paper is to
 find out the properties of this function.

 However, we face two problems in our investigation.
 The first difficulty is that, unlike classical information theory \cite{Cover}
 which is a mature field, a quantum
 theory of information is still being developed. For example,
 despite much past effort,
 a most basic problem in quantum information theory---the classical
 information-carrying capacity of non-orthogonal quantum signals---
 is still generally unsolved \cite{Fuchs:94}.
 Our inability to answer this basic
 question makes the issue of
 secrecy capacity even more intractable. The second problem is the
 fact that,
 even within the context of classical information theory,
 no simple expression has been found for the secrecy capacity
 in general \cite{Maurer:93}. Only lower and upper bounds have been obtained.
 Going to the quantum regime will almost certainly not make things any easier.
 Despite these two difficulties, we shall see in this paper
 that much about the secrecy capacity can still be learned.

 The organization of this paper is as follows. In Section~\ref{sec:coh},
 we introduce a simple spherical symmetric
 EPR-based cryptographic scheme as a toy model and establish
 its security against eavesdropping even in the presence of noise.
 By deriving a lower bound to its security
 capacity, we demonstrate that as the error rate
 tends to zero,  the performance of such a noisy quantum channel
 approaches that of a noiseless one. In Section~\ref{general},
 we generalize our toy model results to more realistic cryptographic
 schemes. First, the assumption of spherical symmetry can
 be relaxed and any choice of two or more non-orthogonal
 measurement bases
 suffices to guarantee the security of
 an EPR-based scheme (provided that the error rate is sufficiently
 small). Second, we show that polarization based cryptographic
 schemes are conceptually equivalent to EPR-based schemes.
 Hence, the proof of the security of quantum cryptography
 and the discussion about secrecy capacity for our toy model trivially
 carry over to polarization based schemes. One interesting implication
 of our results is that one can essentially
 double the efficiency of a most well known cryptographic
 scheme proposed by Bennett and Brassard simply by assigning
 vastly different probabilities to the two conjugate bases.

\section{Coherent Measurements}
\label{sec:coh}
In this Section, we shall establish the security of the following
cryptographic scheme against any attack
even in the presence of noise and investigate its secrecy capacity.
Consider two legitimate
users, traditionally called Alice and Bob who make use of $N$ EPR pairs
to transmit secret messages.
(Here $N$ is supposed to be large.) We assume that there is also
another public, unjammable, classical communication channel between them.
That is, anyone including the eavesdropper can listen to the signals
without worrying about being detected. However, alterations of the
signals are forbidden.
The procedure goes as follows:
Alice prepares $N$ EPR pairs and sends one member of each pair to Bob,
keeping the other member. After receiving {\it all} $N$
transmitted particles, Bob publicly acknowledges his reception.
For each particle that she has kept, Alice chooses a random axis independently
to measure its spin.
Afterwards, she publicly announces the axes that she has chosen for her
measurements,
but not the results.
Bob then performs a measurement on the spin of the other member
of the pair along the axis chosen by Alice.
Ideally, the combined state of
each pair should be a singlet. Thus, the measurement results of Alice
and Bob should be antiparallel. Of course, errors are inevitable due to the
presence of noise in the communication channel.
Nonetheless, most of the spin measurements for the two members of the
various pairs should remain
antiparallel. (Of course, errors may also occur due to the measurement
process itself. For example, a misalignment of the measurement
bases used by Alice and Bob will lead to an increased error rate. However,
in this paper we will not consider this type of errors.)

A more serious problem is the following:
it is conceivable that some of the errors are due to an eavesdropping
attempt by an eavesdropper, traditionally called Eve.
To estimate the extent of eavesdropping,
Alice and Bob may choose randomly a subset of $m$
pairs and declare their measurement results in public. By doing
so, they can compute the error rate for the $m$ test pairs.
If the error rate is found to be unreasonably large,
they assume that eavesdropping has occurred. Thus,
they should reject the whole run and go through the procedure again.
Otherwise, they assume that no successful eavesdropping attempt
has been made. Now they share the remaining $N-m$ bits which may well be
corrupted by the noise and Eve's wiretapping attempt.
So, the hope is
that, at sufficiently small error rates, Alice and Bob can
use well-known schemes of error correction and privacy
amplification to distill out a shorter key of absolute security against
Eve's attack.

The question that we
would like to answer is the following:
For a noisy quantum communication channel,
will Eve be able to obtain a large amount of the
information shared between Alice and Bob
without exposing her eavesdropping attempt by
{\it coherently} interacting the $N$ transmitted particles with an
ancilla? This is in fact the most general eavesdropping strategy.

Before going into the question of coherent manipulation, let us
consider a single EPR pair.
The Hilbert space of an EPR pair is spanned by the singlet $| \psi_0 \rangle $
and the three other states $| \psi_1 \rangle $, $| \psi_2 \rangle $,
and $| \psi_3 \rangle $. Only the singlet state is guaranteed to
give the desirable antiparallel result for the measurement along
any axis chosen by Alice and Bob.
For a noisy channel, the output will generally be a mixed state which
may be described by a density matrix $M$.
One can define the {\it fidelity}
as $F =\langle \psi_0 | M | \psi_0 \rangle $ \cite{Sch}.
It is the probability of the mixed state for passing a test
for being a singlet state. (Thus, $ 0 \leq F \leq 1$).
Being so, it is invariant under simultaneous rotations of the two
particles.
Given an ensemble of identical pairs each described by $M$, one
can estimate its fidelity by the following process. For
each pair, Alice picks a random axis to measure the spin of a member of
the pair.
Bob performs a similar measurement along the same axis on the other member.
Notice that for a given mixed state of an EPR pair described by $M$
and a random axis of measurement chosen by Alice,
the probability that Alice
and Bob's measurements give antiparallel results is $(1 + 2 F) /3$.
The physical reason is simple. If the two members of the $i$-th
pair are
measured along the $z$-axis and found to be antiparallel,
a state in the subspace spanned by the singlet state and
$(|10 \rangle_z + | 01 \rangle_z) / \sqrt{2}$ is consistent
with this result but those in orthogonal complement of this subspace
are not.
If they are measured along the $x$-axis instead, an antiparallel
result will only be consistent with a state in the subspace
spanned by the singlet and $(|10 \rangle_x + | 01 \rangle_x ) / \sqrt{2}$
but not with a state in its orthogonal complement. As we are considering
a random axis, there is spherical symmetry.
Since only one of the three non-singlet states will
give an antiparallel result, we have $P ({\rm antiparallel})=
F + [ (1-F)/3]= (1 + 2 F) /3$.
Intuitively, this means that when an EPR pair is in a non-singlet state,
there is a probability of $2/3$ of failing to give an antiparallel
result. As discussed before, Alice and Bob estimate the error rate
of the channel by publicly announcing the results of
their measurements for $m$ pairs.
For a communication channel with a small error rate, it
would therefore be unwise for Eve to cheat by substituting
non-singlet EPR pairs into the communication channel. Any
amount of substitution with the number of non-singlets
higher than $3/2$ of the original error
rate of the channel is highly likely to lead to an abnormally high error
rate in the $m$ test bits and consequently detection by Alice and Bob.
The curious fact is that, in what follows
this simple observation will play a crucial role in our argument for
the case of coherent manipulation.

\subsection{Security of our EPR Based Scheme\label{sec:EPR}}
To prove the security of the above EPR based scheme, first note
that the most favorable scenario for an eavesdropper Eve
would be to allow her to prepare the states for the $N$ EPR pairs.
Any (more realistic) situation will involve environmental noises
and can be regarded as a special case in which Eve does not
utilize the full control she has on the EPR states.
The most general
state that Eve can prepare is of the form
\begin{equation}
\sum a_{i_1 i_2 \dots i_N} | \psi_{i_1} \rangle | \psi_{i_2} \rangle \dots
 | \psi_{i_N} \rangle  | R_{i_1 i_2 \dots i_N} \rangle ,
\label{EvePre}
\end{equation}
 where $| R_{i_1 i_2 \dots i_N} \rangle $ is the state of the ancilla.
$ | R_{i_1 i_2 \dots i_N} \rangle $ are normalized but need not be
orthogonal to one another.

Suppose Eve is eavesdropping an ideal channel. Alice
and Bob may draw $m$ pairs randomly out of the $N$ transmitted
pairs and publicly compare
their measurement results. They will
regard the transmission of $N$ particles as
untampered only if all the $m$ drawn pairs
show antiparallel results in their measurements. The only
way for Eve to guarantee this is to have all the $N$ pairs in the singlet
state. Therefore, Eve must set
all $a_{i_1 i_2 \dots i_N}$ to be zero except for one state (the tensor product
of singlets). Thus, she will not be able to obtain any information
about an ideal channel.

What about a noisy channel? Suppose,
based on previous communication experience, Alice and Bob
know that the actual channel error rate in the absence of
eavesdropping is $\epsilon$. (Except for Subsection~\ref{secrecy},
we will only
be interested in the regime $\epsilon \ll 1$ in this paper.)
They again draw $m$ pairs randomly from the $N$ transmitted pairs.
We assume $m \ll N $ but $m$ is still large enough for an accurate
estimation of the error rate. In the limit $N \to \infty$, we let
$m \to \infty $ but $m/N \to 0$.
Alice and Bob
may agree that the channel error rate is acceptable if and only if
the number of errors found is in the region say
$[( \epsilon - c \epsilon^2)m ,( \epsilon + c \epsilon^2)m ]$ where $c
={\rm O} (1)$.

The key observation is that most basis vectors in Eq.~(\ref{EvePre})
are highly {\it unlikely} to
give an error rate in this region.
Even if we are generous enough to extend the acceptable
error range to $[0, ( \epsilon + c \epsilon^2)m]$, our conclusion does
not change. Consider a vector of the form $| \psi_{i_1} \rangle
| \psi_{i_2} \rangle \cdots | \psi_{i_N} \rangle$ where $Na$ of the $i_j$'s
(for $j=1, 2, \cdots N$) are nonzero (i.e., non-singlet).
Since the measurement axes are chosen randomly for the $m$ test samples,
such a state on average gives a parallel (i.e., incorrect) result
for $2ma/3$ pairs which is
much larger than the maximal tolerable
number $( \epsilon + c \epsilon^2)m $ for
say $a \geq 2 \epsilon > \epsilon + c \epsilon^2$. Since we assume
$\epsilon \ll 1$,
most of the basis
vectors in Eq.~(\ref{EvePre}) contain far more than $2 N \epsilon $
non-singlet states in a tensor product decomposition with respect to each
particle and tend to give abnormally high error rates.
Therefore, inspired by Shannon \cite{Shannon:48}, we
divide up the Hilbert space of the $N$ pairs into a `typical' subspace
and its orthogonal complement, an `atypical' subspace.
A {\it typical}
subspace is one whose states have exponentially small
probabilities to give an acceptable
error rate \cite{Cover}. A vector in an atypical subspace may fare better.
An example of
a typical subspace $\Lambda$ may be spanned by vectors of the form
$| \psi_{i_1} \rangle | \psi_{i_2} \rangle \dots
 | \psi_{i_N} \rangle$ where the number of non-singlet
$i_j$'s (i.e., $i_j \not= 0$) (here, $ j=1, 2, \cdots N$)
are larger than or equal to
$2 N \epsilon$.
Its orthogonal complement is the atypical subspace. It is spanned by
vectors of the form $| \psi_{i_1} \rangle | \psi_{i_2} \rangle \dots
 | \psi_{i_N} \rangle$ where the number of $i_j$'s ($ j=1, 2, \cdots N$)
that are non-singlet (i.e., $i_j \not= 0$) are less than $2 N \epsilon$.
Notice that, given a state, the number of $i_j$'s that are non-zero
has an invariant meaning.
We shall only consider simultaneous rotations of the two particles
in each pair.
Under arbitrary and independent
rotations of all {\it pairs}, such a state
transforms into a linear superposition
of states with the same number of non-zero $i_j$'s.

Here comes another important observation: the atypical subspace has a small
dimension (as compared to $2^N$, the dimension which gives the $N$
classical bit of information shared between Alice and Bob).
To be more precise, one can give the following generous bound to the
dimension of the atypical subspace
\begin{eqnarray}
 {\rm dim}_{\rm atypical} & \leq & \sum_{a, b, c =0}^{ 2 N \epsilon - 1}
     \left( \begin{array}{c}
            N \\ a
            \end{array} \right)
 \left( \begin{array}{c}
            N -a \\ b
            \end{array} \right)
 \left( \begin{array}{c}
            N -a -b \\ c
            \end{array} \right) \nonumber \\
             &\leq &
  (2 N \epsilon)^3
\left( \begin{array}{c}
            N \\ 2 N \epsilon
            \end{array} \right)
 \left( \begin{array}{c}
            N \\ 2 N \epsilon
            \end{array} \right)
 \left( \begin{array}{c}
            N  \\ 2 N \epsilon
            \end{array} \right) \nonumber \\
             &\leq&
 (2 N \epsilon)^3 2^{3 N H ( 2 \epsilon)} \nonumber \\
             &<& 2^{N [6 H (\epsilon)+ \mu] } \nonumber \\
             &<& 2^{ - N k \epsilon \log_2 \epsilon} ,
\label{dimension}
\end{eqnarray}
where $k$ is a positive constant, $\mu $ a small number of order $\log N /N$,
and $H(x) = - [ x \log_2 x  + (1 -x) \log_2 (1-x)]$  is
the entropy function.
Note that the inequality \cite{Cover}
\begin{equation}
\left( \begin{array}{c}
  N \\ r
\end{array} \right) \leq 2^{N H(r/N)}
\end{equation}
and the concavity of $H(x)$ have been used in the
third and the fourth lines of Eq.~(\ref{dimension}) respectively.
For our purposes, Eq.~(\ref{dimension}) is good enough because
$2^{ - N k \epsilon \log_2 \epsilon}$ is clearly
exponentially smaller than $2^N$,
the dimension that gives $N$ bits of information. Nonetheless,
we remark on passing that a much more refined bound could be found.

Suppose Eve prepare the state
\begin{equation}
\sum_i a_i |{\rm typical}_i \rangle | R_i \rangle + \sum_{j}
  b_j |{\rm atypical}_j
 \rangle |R_j \rangle
\end{equation}
for the combined system of $2N$ particles and ancilla.
Alice and Bob will only accept a run of $N$ pairs if $m$ randomly
chosen samples give a reasonable error rate.
If we average over all the random axes, the probability of
passing such a test
\begin{equation}
P({\rm passing}) \leq
\sum_i |a_i|^2 \exp \left( - f(m) \right) + \sum_j |b_j|^2  ,
\label{passing}
\end{equation}
where $f(m)$ the minimal exponential suppression factor for vectors
in the typical subspace to pass such a test \cite{Cover}.
Notice that $f(m) \to \infty$ as $m \to \infty$.
Thus, the contribution from the typical subspace is bounded
above by $\exp \left( - f(m) \right)$ which goes to $0$ as $m \to \infty$.

\subsection{Eve's Dilemma}
Now the dilemma that Eve faces is clear. In order to have
even just an exponentially
small probability $\exp \left( - f(m)/2  \right)$ of passing
the sample testing, the contribution from the atypical subspace
must exponentially dominate that from the typical subspace.
Without much loss of generality, one can assume that the whole typical
space simply drops out whenever the testing of the $m$ samples
is passed. Therefore, effectively, the dimension of the Hilbert
space is reduced to that of the atypical space.
But the atypical subspace has a small number of dimension
and is incapable of giving Eve much information.

In case the above discussion is still not transparent,
in this paragraph, we show how this selection effect comes about
in more detail. Let us specify the measurement axes for the
$m$ test samples. An outcome is the results (up or down) of
the $2m$ measurements made by Alice and Bob.
Suppose
the initial state of the combined ancilla-particles system is given
by $|u_0 \rangle$. According to the
conventional interpretation of quantum mechanics, if a measurement
gives a outcome $j$ (a state $| v_j \rangle $ for the $m$ test pairs),
the state of the system will be projected onto $|j \rangle=
\left( | v_j \rangle \langle v_j|
\otimes \openone_{\rm other}
 \right) |u_0 \rangle$. where $\openone_{\rm other}$ is the identity
operator for the other degrees of freedom (i.e., the $N-m$ remaining
pairs and the ancilla).
The probability $p_j$ of this outcome $j$ occuring is given by
$ \sum_i \left| \left( \langle v_j| \langle r_i | \right)
 |u_0 \rangle \right|^2 $ where
$\langle r_i|$ denotes the state of the other degrees of freedom and
the sum is over a complete basis. Since Alice and Bob will reject all
measurement results with abnormally high
error rates, most outcomes $j$ will be rejected.
Under the assumption
that the $m$ samples pass the test,
the state of the combined system after the test
will be described by a density matrix
\begin{equation}
\rho_c = {\sum_j}' (p_j / {\sum_l}' p_l) |j \rangle
\langle j|,
\label{pass}
\end{equation}
where the sums are over those outcomes that pass the test.
The crucial insight is, however, that all basis vectors
are not created equal. As noted before, if we consider a tensor
product state of $n$ non-singlets and $N- n$ singlets, under arbitrary
and independent rotations of all pairs, it will transform into
a linear superposition of states that are also made of $n$ non-singlets
and $N-n$ singlets. (Here, particles in the same
pair are only allowed to be rotated by the same amount because we are only
interested in measurements that are done along the same axis on the two
members of a pair.)
The likelihood of a state in passing the test
depends on $n$.
Vectors in the typical space have a large $n$ and are
exponentially unlikely (as a function of $m$) to pass the test
while those in atypical space may fare better.
Therefore, any realistic chance of passing the test is due to the
atypical space (which consists of vectors
of small $n$). This selection effect effectively eliminates
the whole typical space from our consideration.

Moreover, the atypical space has a small
dimension $2^{-N k \epsilon \log \epsilon}$
as given by Eq.~(\ref{dimension}).
An upper bound on the amount of information that Eve can acquire
by measuring the ancilla is given by the Holevo's theorem \cite{Holevo:79}:
\begin{equation}
I^{\rm eve}_{\rm max}= S(\rho_R) = - {\rm Tr} \rho_R \log \rho_R ,
\end{equation}
where
\begin{equation}
\rho_R = {\rm Tr}_{\rm particles}  \rho_c =
 {\rm Tr}_{\rm particles} {\sum_j}' (p_j / {\sum_l}' p_l)
| j \rangle \langle  j|,
\end{equation}
is the reduced density matrix for the ancilla given that the $m$ samples
pass the test.
We obtain an upper bound
\begin{equation}
I_{\rm max}^{\rm eve} \leq  N [-k \epsilon \log \epsilon + \theta],
\end{equation}
where $\theta$ is a small correction term coming form the typical space.
Asymptotically, $\theta$ can be made as small as one is pleased
by taking $m \to \infty$.

Eve may well have some a priori information about the measurements.
The point is that the probability
of getting an ``up'' in Alice's (or Bob's) measurement
may well depend on the orientation of the axis chosen. The probability
that the spin measurements by Alice and Bob are antiparallel can
also have such an orientation dependence. Thus, the
mutual information shared by Alice and Bob may actually be smaller
than $N$ bits.
Nevertheless, any correction term must be of the order $ -N \epsilon
\log \epsilon$.
For sufficiently small error rate $\epsilon$,
the secrecy capacity $C_s$ of the channel (per EPR pair) therefore satisfies
\begin{equation}
C_s >  [ 1 - k' \epsilon \log \epsilon] ,
\end{equation}
where $k'$ is some constant.
Notice that as the fidelity $F \to 1$, $\epsilon \to 0$ and $C_s \to  1$.
Therefore, an arbitrarily small error rate implies a secrecy capacity
arbitrarily close to the ideal channel capacity (which is
one bit per EPR pair). Notice that Eve can still mess up
the results of say
${\rm O} (\log N) $ pairs without worrying about being detected
because the portion of pairs tested
$m/N \to 0$. However, this has no effect on the secrecy capacity.
What we have shown is that any attempt to obtain ${\rm O}(N)$
bits of information
{\it will} be almost surely detected.

What is the principle underlying the
security
of an EPR-based cryptographic scheme? Ekert \cite{Ekert:91}
suggested that it comes Bell's theorem.
However, Bennett, Brassard and Mermin \cite{Bennett:92} later
proved the security of EPR-based schemes
without invoking the Bell's theorem.
Nevertheless, both works only addressed noiseless channels.
Here, we would like to propose an alternative viewpoint which
remains useful even for noisy channels.
{}From an information-theoretic point of view, the security of an EPR-based
quantum cryptographic scheme can be traced back to the observation
that in quantum physics,
knowing completely the state of a composite system {\em does not}
guarantee complete knowledge of the states of the individual constituents
because of the presence of entanglement entropy.
For instance, the entropy of each member of a perfect EPR pair is non-zero
even though the total entropy of the pair is zero. Consequently,
two observers are able to use an EPR pair to transmit a random
but secret bit of classical information. Heuristically, in the presence
of noise, we expect that transmission of secret information
is still possible as long as the ``entanglement entropy''
remains larger than the entropy of the composite system.

\section{Generalizations}
\label{general}
For simplicity, we have discussed only the case in which Alice chooses
the bases for her measurement randomly.
Some generalizations of the above result are possible.
A moment of thought will convince the reader that a similar
proof can be formulated to the case when say only two
non-orthogonal bases are used.

Let us consider another modification of the procedure.
What if Alice performs
the measurement on the particles in her share
{\it before} sending Bob the other member of the EPR pair?
Since the operators Alice uses in the measuring process
only act on the particles
in her share, they must commute with
any operators (which may be used by Eve and Bob)
that act on the particles in Bob's control.
It is, thus, immaterial\footnote{Incidentally, a
related idea is used in the context of quantum
computing in a recent preprint by
Griffiths and Niu \cite{Semiclassical}.} whether Alice performs the
measurement first and sends out the rest second or the other way
round {\it provided} Alice announces her basis only {\it after} Bob's
public acknowledgment of his reception of {\it all} $N$ particles in his
share.
Since all particles are already in Bob's hand, it is too late for
Eve to do anything.
Notice that if
Alice announces her basis too early (say she announces
her basis each time Bob acknowledges his reception of one particle),
this argument does {\it not} preclude
an intelligent eavesdropper from obtaining a substantial amount of
information about Alice and Bob's
measurement results.

\subsection{Polarization Based Schemes}
Remarkably, the above simple observation---that it is immaterial
whether Alice measures the spins
first and sends out the other particles second or vice versa---
has deep consequences.
So far our discussion has been concentrated on EPR based cryptographic
schemes. However, another class of schemes that are based on
the polarization of photons has been discussed in the literature.
For instance, in 1984 Bennett and Brassard \cite{BB:84}
proposed a scheme (BB84) in
which the key distribution between Alice and Bob is done
by sending photons over optical fiber. To detect
eavesdropping, Alice chooses randomly with {\it equal}
probability between
the rectilinear basis (i.e.,
horizontal and vertical) and the diagonal basis ($45{^\circ} $ and
$135^{\circ}$).
A horizontally polarized photon can represent a $0$ and vertical a $1$.
Similarly, a $45^{\circ}$ polarized photon can represent a $0$ and
$135^\circ$ a $1$. $0$'s  and $1$'s are chosen with
equal probability. Alice then transmits a photon in the basis
of her choice. Similarly, Bob performs a measurement along
the rectilinear basis and the diagonal basis with equal
probability. Afterwards, both Alice and Bob publicly announce
the bases that they have chosen, but not the results of their
measurements. Their bases will therefore agree with each other only
half of the time. As in the case of EPR based schemes, they can
then choose a subset of
those measurements that are done in the same bases and compare
the results in public. From the error rate of the $m$
test samples, they can
estimate the error rate for the whole run and hence the degree
of eavesdropping. They can then decide whether to accept the
run or to reject the run and do it again.

As argued by Bennett, Brassard and Mermin \cite{Bennett:92}, the
two classes of schemes (EPR based and polarization
based) are conceptually equivalent. The point is that
Alice could have prepared each photon by producing an
EPR pair of photons and measuring
one member along a random axes (rectilinear or
diagonal), letting the other
particle, now in a known random one of the four states, pass to Bob.
We remark that this argument remains valid even for a noisy channel.

There is still, however, one minor difference.
So far we have assumed that, after the transmission of
the $N$ EPR pairs, Alice informs Bob
of her basis of measurement for her particle in each pair
and Bob is supposed to measure the spin of the corresponding partner along
the {\it same} basis.
It is of course experimentally difficult for Bob to store up
a large number of photons to wait for Alice's announcement of
her bases. Allowing Alice and Bob to choose
between the two bases (rectilinear and diagonal)
{\it independently} as in the BB84 scheme is definitely
more realistic. Is this going to affect our conclusions?

The answer is no. Recall that all we need in our proof is
to use a small subset of the EPR pair to estimate the
error rate when the axes Alice and Bob used {\it do} agree.
Whether the axes for other measurements agree or not is irrelevant.
Our proof of security of
EPR based cryptography, therefore, automatically implies
the security of polarization based schemes even if
Alice and Bob choose their measurement axes independently.

There are two alternative points of view regarding the underlying principles
governing the security of quantum cryptography.
The first and more well publicized
point of view, which has been discussed in Section~\ref{sec:intro},
is that measurements performed
on non-orthogonal states in quantum mechanics
generally lead to disturbance. For a noiseless channel,
it leads to
the generalized ``no-cloning'' theorem (see the Appendix).
In our opinion, the trade-off between information gain
and disturbance by an eavesdropper in a {\it noisy} channel
remains to be studied in more detail.
The second point of view, which
has been discussed in
the last paragraph of Section \ref{sec:coh}, is that
the security of
quantum cryptography lies on the possibility in quantum mechanics
of the ``entanglement
entropy'' between two subsystems being larger than the entropy of
the whole system.
The equivalence between EPR and polarization based schemes
suggests that these two alternative points of view
are in fact equivalent.

\subsection{Doubling Efficiency in BB84}
In the BB84 scheme, Alice and Bob choose their measurement
axes from two conjugate bases (rectilinear and diagonal)
independently and with {\it equal} probability.
A drawback of such a scheme is the reduction of the ideal
secrecy capacity to half of the optimal value (i.e., $1/2$ bit per
pair vs $1$ bit per pair because only half of time will the two
independently chosen bases by Alice and Bob agree).
However, we would like to remark that
the restriction of equal probability in choosing the two
bases is totally redundant. Conceptually, the probability of
choosing the rectilinear basis can be made much larger than
the probability for the diagonal basis. At small error rates,
this will lead to a secrecy capacity
which is almost {\it double}
of the original BB84 scheme.

More explicitly,
given a noisy channel with error rate $\epsilon \ll 1$. Suppose
they choose the rectilinear and diagonal bases with probabilities
$1 - \omega$ and $\omega$ respectively
(where $\omega \ll 1$). For the transmission of $N$ photons, there
are on average $N \omega^2$ photons for which both Alice and Bob measure
along the diagonal basis. They can for example publicly compare their
measurement results for those $N \omega^2$ photons. In addition, they
also randomly choose $N \omega^2$ photons from the set for
which they both measure along the rectilinear axis. They can decide that
the error rate is acceptable if and only if it is less than $ 2 \epsilon$.
Now given any $\omega \ll 1$, there exists an $N_0$ such that,
for the transmission of $N > N_0 $ photons, any eavesdropping attempt
to get more than ${\rm O}(- N \epsilon \log \epsilon)$ bits of
information about the state of the transmitted particles will almost
surely be detected. Thus, this scheme with different probabilities
for the two bases is clearly secure. Furthermore, it has the
benefit that, in the limit
$\omega \to 0$, we obtain essentially double of the efficiency of
the scheme proposed by Bennett and Brassard.
It is of practical interest to investigate whether this observation
will lead to the design of more efficient protocols for
say quantum oblivious transfer and quantum bit commitment \cite{ot}.

\subsection{Other Generalizations}
Of course, in practical applications, the quantum signals used in
BB84 are low-intensity light pulses rather than
ideal single photon pulses. In that case, we must consider the possibility
of beamsplitting attack. We shall, however, pursue this problem
no further in this paper.

There are other protocols of polarization based cryptographic schemes.
For instance, two rather than four non-orthogonal quantum states
are used in a scheme proposed by Bennett \cite{Bennett}.
We believe that the techniques developed in this
paper can be used to prove the security of this kind of schemes as well.

\subsection{Properties of Secrecy Capacity\label{secrecy}}
Let us return to the subject of secrecy capacity.
So far, we have only derived a lower bound to the secrecy capacity of
a quantum communication channel. Can we derive an upper bound?
At first sight, the answer is a simple yes:
One can just choose an eavesdropping strategy and
compute the amount of information acquired through it.
On second thought, it is not so simple. Given an eavesdropping
strategy, Alice and Bob may counteract
by changing their procedure. Instead of measuring the state of
each carrier of quantum signal and perform {\it classical} error correction
and privacy amplification as assumed before, they may manipulate the
state of a number of the particles {\it coherently} \cite{Jozsa:un}. It is not
entirely inconceivable that such {\it quantum} processing of
signals might give the users more information than any classical methods.
This subject deserves further investigations.

Baring coherent manipulation by the users, one can show that
the secrecy capacity is a convex function of the error rate.
In other words, $C_s ( a x+ by) \leq a C_s(x) + b C_s(y)$ for all $a, b \leq
1$ such that $a+b =1$. The idea is the following:
given strategies $S_x$ and $S_y$ that correspond to
the preparation of the
ancilla-particles states $| u_x \rangle$ and $|u_y \rangle$ with
error rates $x$ and $y$ respectively, Eve can construct the tensor
product state
$|u_x \rangle \otimes |u_x \rangle \otimes \cdots  \otimes |u_x \rangle
\otimes |u_y \rangle \otimes |u_y \rangle  \otimes \cdots  \otimes
|u_y \rangle $
(with arbitrary numbers of $|u_x \rangle$'s
 and $ |u_y \rangle $'s and permutations of the particles involved
if desired) to give an error rate $ax+by$. If the legitimate users
knew about the decomposition of the channel, the secrecy capacity
they could achieve would be $a C_s(x) + b C_s(y)$. Their ignorance
certainly makes things worse. Hence,
$C_s ( a x+ by) \leq a C_s(x) + b C_s(y)$.

Another interesting question is the following: For the EPR
based scheme discussed in section~\ref{sec:coh}, what is the minimal
value of $\epsilon$ (call it $\epsilon_{\rm min}$)
such that $C_s(\epsilon) = 0$? That is to
say the channel is too noisy to be of any use. Clearly, $C_s (1/2) = 0$
(because $P(\mbox{parallel}) = P(\mbox{antiparallel}) = 1/2$)
and thus $\epsilon_{\rm min}  \leq 1/2$.
The value of $\epsilon_{\rm min}$ is an interesting open question.

The EPR scheme introduced in Section~\ref{sec:coh} is designed to be
spherical symmetric so that the problem can be characterized by just one
parameter, namely the fidelity or the error rate. This is why the
secrecy capacity is a function of one variable.
In a more general setting, more than one parameters may be needed
for the characterization of the noise level of
a quantum communication channel.
Consequently, the secrecy capacity will be a function of multiple variables.
As far as the legitimate users are concerned, the output of a communication
channel is related to the input by a superscattering matrix. The goal of the
users is to choose their inputs
so as to maximize the information of the output and minimize the information
leakage to the environment at the same time.

\subsection{Conclusions}
We have proved that an EPR based quantum cryptographic scheme is secure against
coherent measurements by eavesdroppers. Our proof relies on the law
of large number. The dimension of the space of states that are consistent
with a small rate is exponentially smaller than the dimension of the
whole Hilbert space, Thus, by testing the error rate for
a small subset of signals, one can effectively eliminate most dimensions.
Consequently, an eavesdropper is unable to get much information.
Moreover, we prove that a polarization based cryptographic scheme is
conceptually equivalent to an EPR based scheme.
Our proof of the security of quantum cryptography therefore carries over
to the former. The secrecy capacity of a quantum channel is also
investigated.

On a conceptual level,
our work suggests that the two alternative points of view
(namely $(1)$ the ``no-cloning'' theorem
for non-orthogonal quantum states and $(2)$ that the entanglement
entropy between subsystems being larger than the
entropy of the whole system) concerning the principles underlying the
security
of quantum cryptography are in fact equivalent.
One practical implication of our results is
that one can double the efficiency
of the cryptographic scheme proposed by Bennett and Brassard \cite{BB:84}
(BB84) simply by assigning vastly different probabilities to the two conjugate
bases.
Finally, we remark that the beamsplitting attack remains to be
addressed in future investigations.

\section{Acknowledgment}
H.-K. L. thanks J. Preskill for introducing him to the subject of
quantum computation and R. Josza for providing references.
We also thank F. Wilczek for critical comments.
This work is supported by DOE grant DE-FG02-90ER40542.
\appendix
\section*{Generalized Quantum ``No-Cloning'' Theorem}
 Suppose we are given a particle that can be in either one of the two
 non-orthogonal states, $| u_1 \rangle$ or $|u_2 \rangle$ of
 a two-dimensional Hilbert space.
 Here, we prove that it is impossible to obtain information
 distinguishing between the two possibilities without perturbing
 its state.
 A simple proof goes as
 follows.
 An eavesdropper may generally couple an ancilla in the state
 $| \Psi \rangle$ to the particle and
 evolve the combined system. To avoid detection, the final state
 of the signal has to remain unchanged. Now suppose
 $ U \left( | u_i \rangle  | \Psi \rangle \right) =| u_i \rangle
 | \Phi_i \rangle $. Since $U$ is unitary,
\begin{equation}
 \langle u_1 | u_2 \rangle =
 \langle \Psi | \langle u_1 | u_2 \rangle
 | \Psi \rangle =  \langle u_1 | u_2 \rangle \langle \Phi_1 | \Phi_2 \rangle .
\end{equation}
 Since  $\langle u_1 | u_2 \rangle \not= 0$, it follows that
 $\langle \Phi_1 | \Phi_2 \rangle =1 $. Thus, $| \Phi_1 \rangle=
 | \Phi_2 \rangle $ and no information can be obtained.

\end{document}